\documentclass{article}

\usepackage{graphicx}
\usepackage{float}
\usepackage[square,sort,comma,numbers]{natbib}
\usepackage{subcaption}			
\usepackage[super]{nth}
\usepackage{amsmath}
\usepackage{hyperref}
\usepackage[dvipsnames]{xcolor}  
\usepackage{lineno} 
\usepackage{placeins}
\usepackage{array}
\usepackage{tabu}
\usepackage{amssymb}
\usepackage{mathtools}
\usepackage{lineno}
\usepackage[ruled,vlined]{algorithm2e}
\usepackage{algorithmic}
\usepackage{wrapfig}
\usepackage{booktabs}       
\usepackage{romannum}
\usepackage{float}
\usepackage{authblk}

\title{Reinforcement Learning for Active Flow Control in Experiments}

\author[1,$\dagger$]{Dixia Fan\thanks{Email address for correspondence: dfan@mit.edu}}
\author[2,$\dagger$]{Liu Yang}
\author[1]{Michael S Triantafyllou}
\author[2]{George Em Karniadakis}
\affil[1]{Department of Mechanical Engineering, Massachusetts Institute Technology, Cambridge, MA 02139, USA}
\affil[2]{Division of Applied Mathematics, Brown University, Providence, RI 02912, USA}
\affil[$\dagger$]{These authors contributed equally to this work.}
\date{\vspace{-5ex}}

\begin{document}
\pagenumbering{arabic} 

\maketitle

\begin{abstract}

We demonstrate experimentally the feasibility of applying reinforcement learning (RL) in flow control problems by automatically discovering active control strategies without any prior knowledge of the flow physics. We consider the turbulent flow past a circular cylinder with the aim of reducing the cylinder drag force or maximizing the power gain efficiency by properly selecting the rotational speed of two small diameter cylinders, parallel to and located downstream of the larger cylinder. Given properly designed rewards and noise reduction techniques, after tens of towing experiments, the RL agent could discover the optimal control strategy, comparable to the optimal static control. While RL has been found to be effective in recent computer flow simulation studies, this is the first time that its effectiveness is demonstrated experimentally, paving the way for exploring new optimal active flow control strategies in complex fluid mechanics applications.
\end{abstract}

\section{Introduction}

The classical paradigm for designing fluid control strategies, for example in the case of reducing the drag force on a bluff body through active control, consists of a first stage when we explore and understand the physics of the problem over a wide parametric range, followed by careful modeling and developing specially designed control strategies to exploit the understanding, and culminating in a optimal tuning of the control parameters \citep{brunton2015closed}. The process involves careful computational or experimental investigation, and intuition obtained through the investigation, resulting in heuristically derived schemes. Hence, this procedure is extremely slow to yield effective results. In recent years, the application of machine learning in fluid control problems has received increasing attention \citep{brunton2019machine}, because it offers a totally different paradigm to arrive quicker at results. The combination of proper machine learning tools with domains of expertise in fluid mechanics could shift the classical paradigm by directly optimizing the control strategy, reducing or even eliminating human involvement in modeling and design of the control strategies.

Among the machine learning tools, reinforcement learning (RL) offers especially intriguing opportunities for a quick progress, as it has demonstrated the capability of achieving ``superhuman'' performances in board games \citep{silver2017mastering}, and a capability for tackling complex, high-dimensional continuous control tasks \citep{haarnoja2018soft}. Recent explorations of RL for fluid mechanics problems include fish bio-locomotion \citep{gazzola2014reinforcement,verma2018efficient}, motion and path planning for aerial/aquatic vehicles \citep{colabrese2017flow, novati2019controlled}, active flow control for bluff bodies \citep{ma2018fluid,rabault2019artificial}, and foil shape optimization \citep{viquerat2019direct}.

\begin{figure}[ht]
    \centering
    \includegraphics[width=0.5\columnwidth,keepaspectratio]{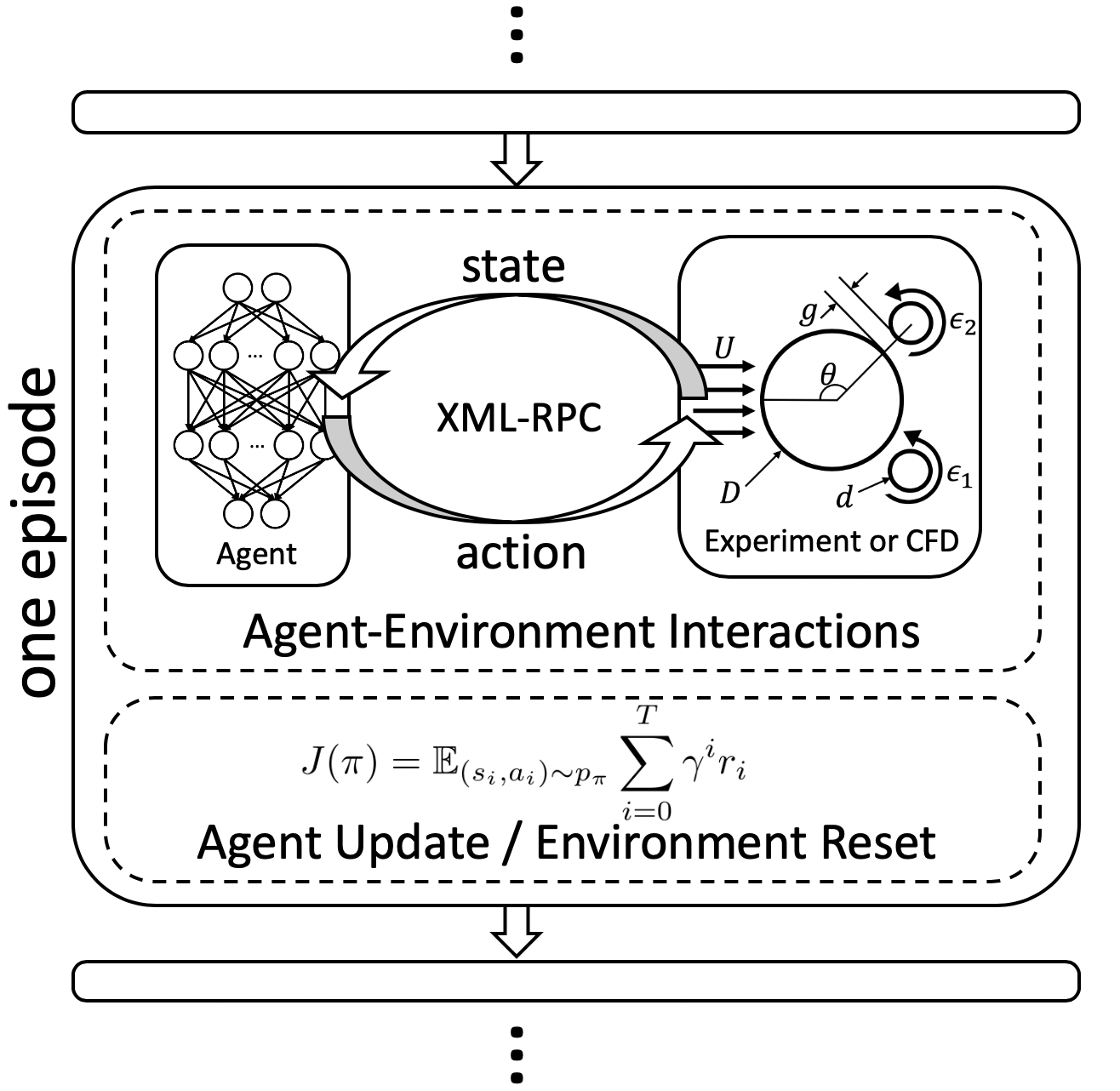}
    \caption{Sketch of the reinforcement learning process in both experimental and simulation environments for one episode. Each episode consists of two stages (the dashed blocks): in the first stage the agent interacts with the experimental or simulation environment via XML-RPC protocol at a fixed frequency, each interaction consisting of a state inquiry and an action decision; in the second stage the agent updates its policy based on the experience collected while waiting for the reset of environment. }
    \label{fig:Sketch2}
\end{figure}

To authors' best knowledge, RL applications in fluid problems are so far limited to only computer simulations. In order to demonstrate the feasibility of applying RL to experimental fluid mechanics, we address experimentally the problem of drag reduction and power gain efficiency maximization in a bluff cylindrical body equipped with two small rotating control cylinders.  A sketch of the model used is shown in Fig. \ref{fig:Sketch2}. The problem has been studied both experimentally \citep{schulmeister2017flow} and numerically \citep{zhu2015simultaneous}, investigating the effects of (a) the small to large cylinder diameter ratio $d/D$, (b) gap ratio $g/D$, (c) the smaller control cylinder configuration, and (d) the rotation rate $\epsilon = \frac{\omega d}{2U}$ on the fluid forces and patterns. Here, $D$ is the diameter of the main cylinder, $d$ is the diameter of the smaller control cylinder, $g$ is the gap between the main and each of the smaller cylinders, $\omega$ is the rotation speed of the smaller cylinders and $U$ is the oncoming velocity. Past results showed that the counter rotating cylinder pair could effectively reduce the main cylinder drag force as well as diminish the oscillatory lift force by suppressing the vortices shed in the wake \citep{beaudoin2006drag}. The physics behind this phenomenon is that when the control cylinders are placed at appropriate locations and rotate at a sufficiently fast speed, they are able to interact with the main cylinders separating boundary layer and cause it to reattach, resulting in a narrower wake behind the cylinder, hence significantly reducing the pressure drag.

\begin{table}
\centering
\begin{tabular}{c|c|c}
    Parameter & Experiment & Simulation\\
    \hline
     $D$ & 5.08cm & 1 \\
     $d/D$ & 0.125 & 0.125 \\
     $g/D$ & 0.025 & 0.025 \\
     $L/D$ & 9 & - \\
     $\theta$ & $2\pi/3$ & $2\pi/3$ \\
     $Re_D$ & 10,160 & 500 \\
     $\delta t$ & 0.1s & 0.12 \\
     $\epsilon^{max}$ & 3.66 & 5\\

\end{tabular}
\caption{Experimental and simulation parameters.}
\label{tab:para}
\end{table}

In this paper, we first describe our active flow control procedure and then present results of applying RL in both an experimental as well as a simulation environment. We demonstrate that with a properly designed reward function and noise reduction, the agent can learn the control strategy that is close to the optimal static control to reduce the system drag or maximize the power gain efficiency.

The rest of the paper is organized as follows: Section \ref{secMethod} presents an overview of the experimental and computational model and procedure as well as the RL algorithm. Section \ref{secResDis} discusses the experimental and simulation results. Section \ref{secConc} summarizes the main findings of the paper.

\section{Methodology Description}
\label{secMethod}

\subsection{Experimental model and procedure}

The sketch in Fig. \ref{fig:Sketch2} outlines the experimental procedure and highlights one episode of the learning process, corresponding to one towing experiment lasting 40 seconds. At the beginning of each experiment, the control cylinders are held still for four seconds to ensure a fully developed wake. Then, the RL agent starts to interact with the environment via a state inquiry and an action decision at 10Hz ($\delta t = 0.1s$). The states in the current experiment are the drag and lift coefficients $C_d$ and $C_l$ on the three cylinders altogether, and can be calculated as follows,
\begin{equation}
\label{eq:CTotal}
    C_d = \frac{F_d}{0.5\rho U^2 DL}, C_l = \frac{F_l}{0.5\rho U^2 DL},
\end{equation}
where $\rho = 1000 kg/m^3$ is the fluid density, and $F_d$ and $F_l$ are the average drag and lift forces over $\delta t$. After completing one experiment, the carriage is brought back to the starting point. Then, the policy of the RL agent is updated based on the experience learnt from all the previous experiments up to that time, while the environment is reset and prepared for the next experiment. A two-minute pause is imposed between towing experiments to avoid cross-contamination of the results between successive experiments. 

The policy of the RL agent in the current work is only updated between experiments, instead of at every agent-environment interaction, to reduce delay of action due to the limitation of our hardware. The control and data collection interface are developed in $C\#$ language, and the RL agent is implemented in Python language based on the deep learning package TensorFlow \citep{abadi2016tensorflow}. The XML-RPC protocol is then applied for data communication between the cross-language platforms, which allows us to take advantage of the machine learning tools developed in Python, as well as the established experimental and computational platforms in other languages with minimum effort.

\subsection{Simulation model}

In addition to the experiment, we also employ a simulation model implemented in the Lily-Pad solver~\citep{weymouth2015lily}.  We conduct two-dimensional numerical simulations and visualize the flow around the main and control cylinders via the same procedures described in the last subsection. 

The simulation resolution is selected to be 24 grids per main cylinder diameter $D$ and a domain size of $8D\times 16D$. The Reynolds number based on the main cylinder is $Re_D = 500$, the same as in the simulation work by \cite{schulmeister2017flow}. Based on $Re_D$, the fixed non-dimensional time step is selected equal to $0.0075$, and the state inquiry and action decision are made every 16 time steps. In each episode, the RL agent starts actions at non-dimensional time $t = 80$ when the wake behind the cylinders have fully developed, and terminates at $t = 130$. Each simulation takes 10 minutes on a single core of a Dell workstation precision tower 5810.

The configuration parameters for the simulation are listed in Table \ref{tab:para}. In the simulation, the states for the RL agent are different from those in the experiment, selected to be the drag and lift coefficient ($C_d^D$ and $C_l^D$) of the main cylinder alone, calculated as follows, 
\begin{equation}
\label{eq:CLarge}
    C_d^D = \frac{F_d^D}{0.5\rho U^2 DL}, C_l^D = \frac{F_l^D}{0.5\rho U^2 DL},
\end{equation}
where $F_d^D$ and $F_l^D$ are the average drag and lift forces on the main cylinder alone over the 16 simulation time steps. 

\subsection{Reinforcement learning}
Reinforcement learning involves an agent interacting with the environment, aiming to learn the policy that maximizes the expected cumulative reward. At each discrete time step $i$, the agent makes an observation of the state $s_i \in \mathcal{S}$, and selects corresponding actions $a_i \in \mathcal{A}$ with respect to the policy $\pi: \mathcal{S}\rightarrow \mathcal{A}$ to interact with the environment, then receives a reward $r_i$. The objective is to find the optimal policy $\pi_{\phi}$ parameterized by $\phi$ which maximizes the expected cumulative reward,
\begin{equation}
    J(\pi) = \mathbb{E}_{(s_i,a_i)\sim p_{\pi}}\sum_{i=0}^T\gamma^i r_i,
\end{equation}
where $\gamma \in (0,1]$ is a discount factor and $p_{\pi}$ denotes the state-action marginals of the trajectory distribution induced by the policy $\pi$.

As mentioned in the previous subsections, in the current work the state is the concatenation of $C_l$ and $C_d$ in experimental environments, or $C^D_l$ and $C^D_d$ in simulation environments. The action is the concatenation of $\epsilon_1/\epsilon^{max}$ and $\epsilon_2/\epsilon^{max}$. The reward received in each time step is induced from the state and action in the subsequent interaction. The detailed formulations of the reward and the comparisons will be presented in the next section.

The update of the agent follows one of the state-of-the-art deep RL algorithms, viz. the Twin Delayed Deep Deterministic policy gradient algorithm (TD3) \citep{fujimoto2018addressing}. In this paper, all the neural networks are feedforward neural networks with 2 hidden layers, each of width 256. The discount factor $\gamma$ is set as 0.99. The policy exploration noise is set as $\mathcal{N}(0,0.1^2)$ in the task of drag reduction, and $\mathcal{N}(0,0.01^2)$ in the task of system power gain efficiency maximization. We use the Adam optimizer with learning rate $10^{-4}$ with batch size $512$, and update the critic networks $1000$ iterations in each episode, while updating the actor and target networks every $2$ iterations. Other hyperparameters are inherited from \cite{fujimoto2018addressing}.

\section{Results and Discussion}
\label{secResDis}

\subsection{Experimental validation for constant rotating control cylinders}
\label{secex}
\begin{figure}
\centering
\begin{subfigure}{0.28\columnwidth}
\centering
\includegraphics[width=\columnwidth,keepaspectratio]{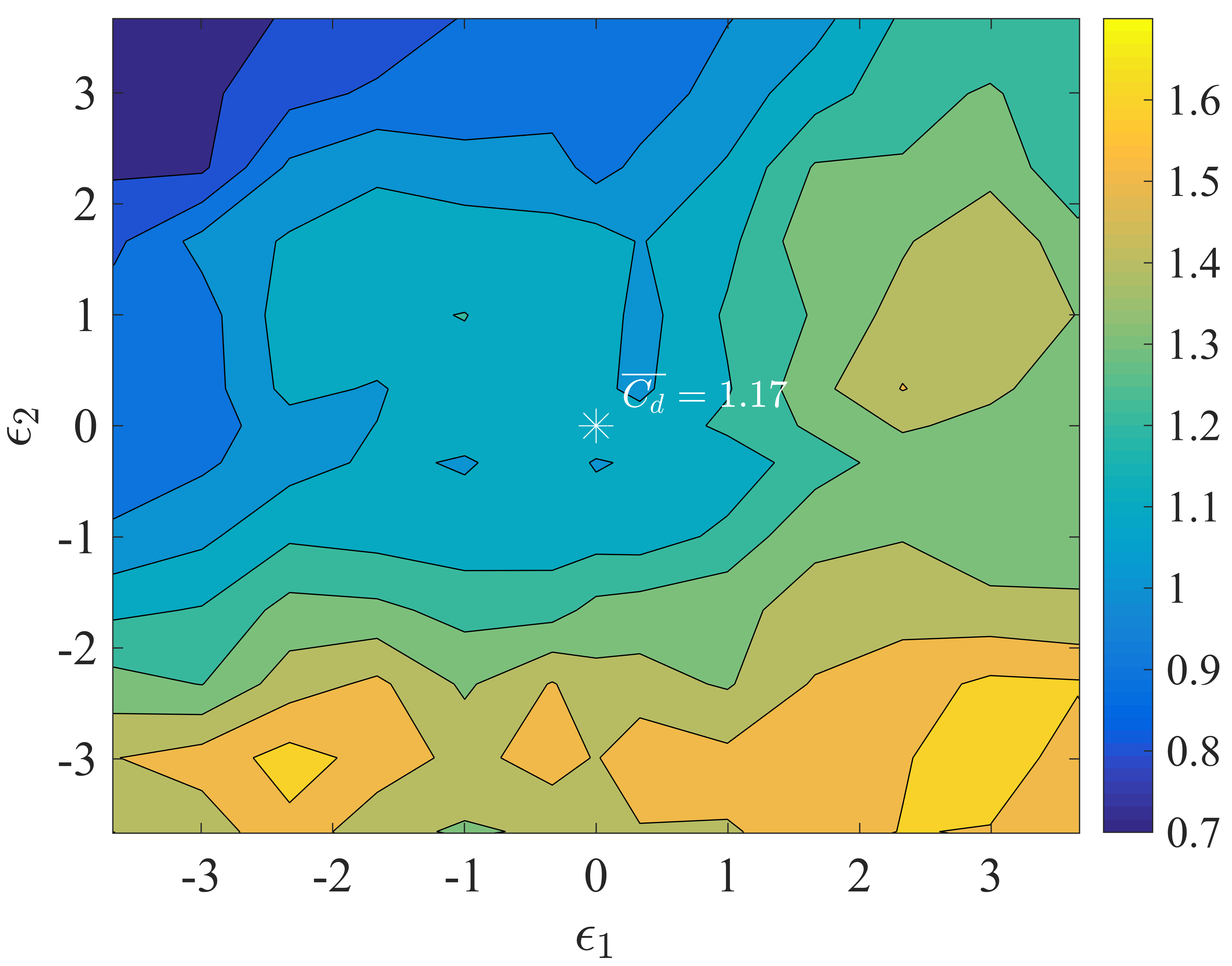}
\subcaption{}
\end{subfigure}
\begin{subfigure}{0.28\columnwidth}
\centering
\includegraphics[width=\columnwidth,keepaspectratio]{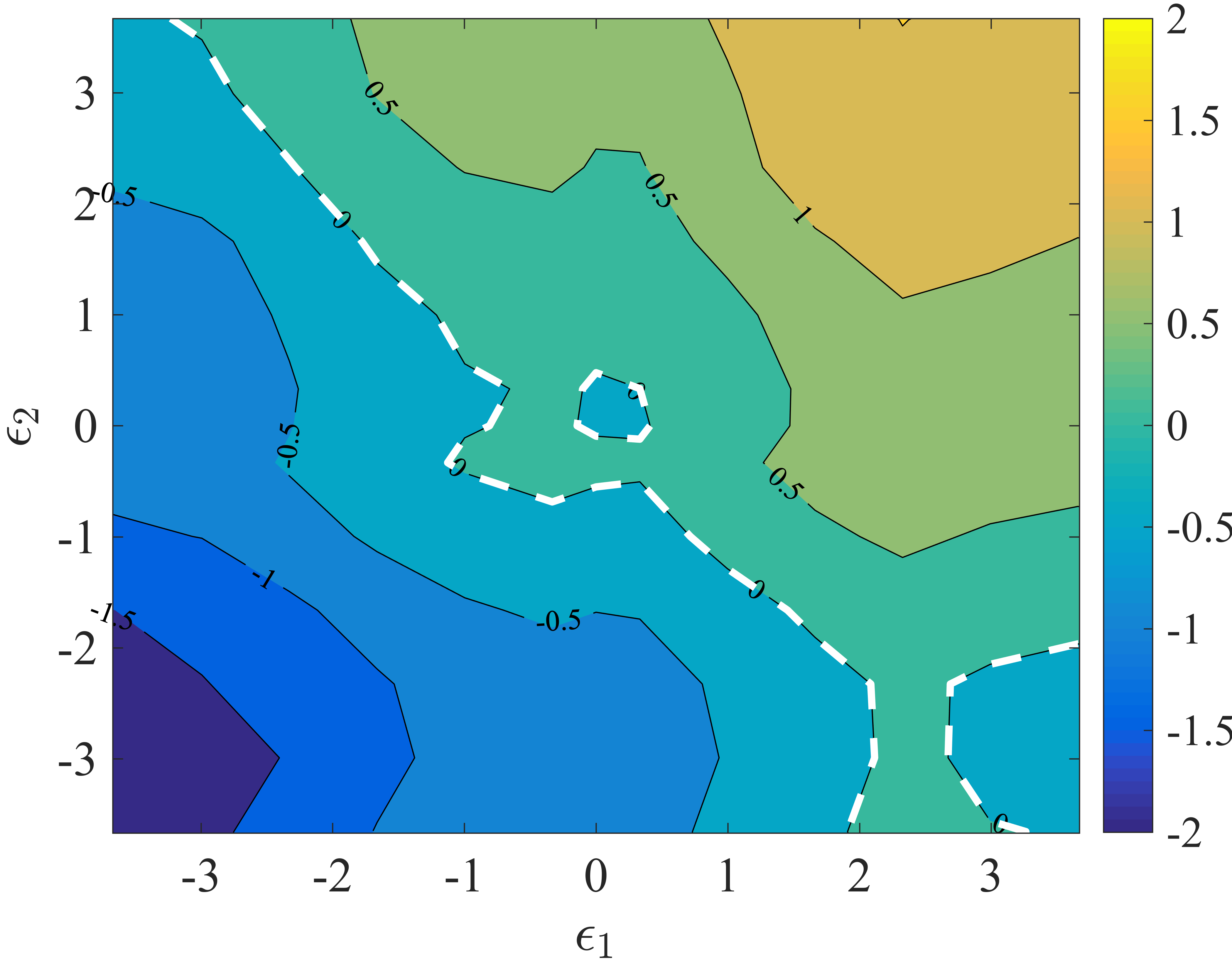}
\subcaption{}
\end{subfigure}
\begin{subfigure}{0.28\columnwidth}
\centering
\includegraphics[width=\columnwidth,keepaspectratio]{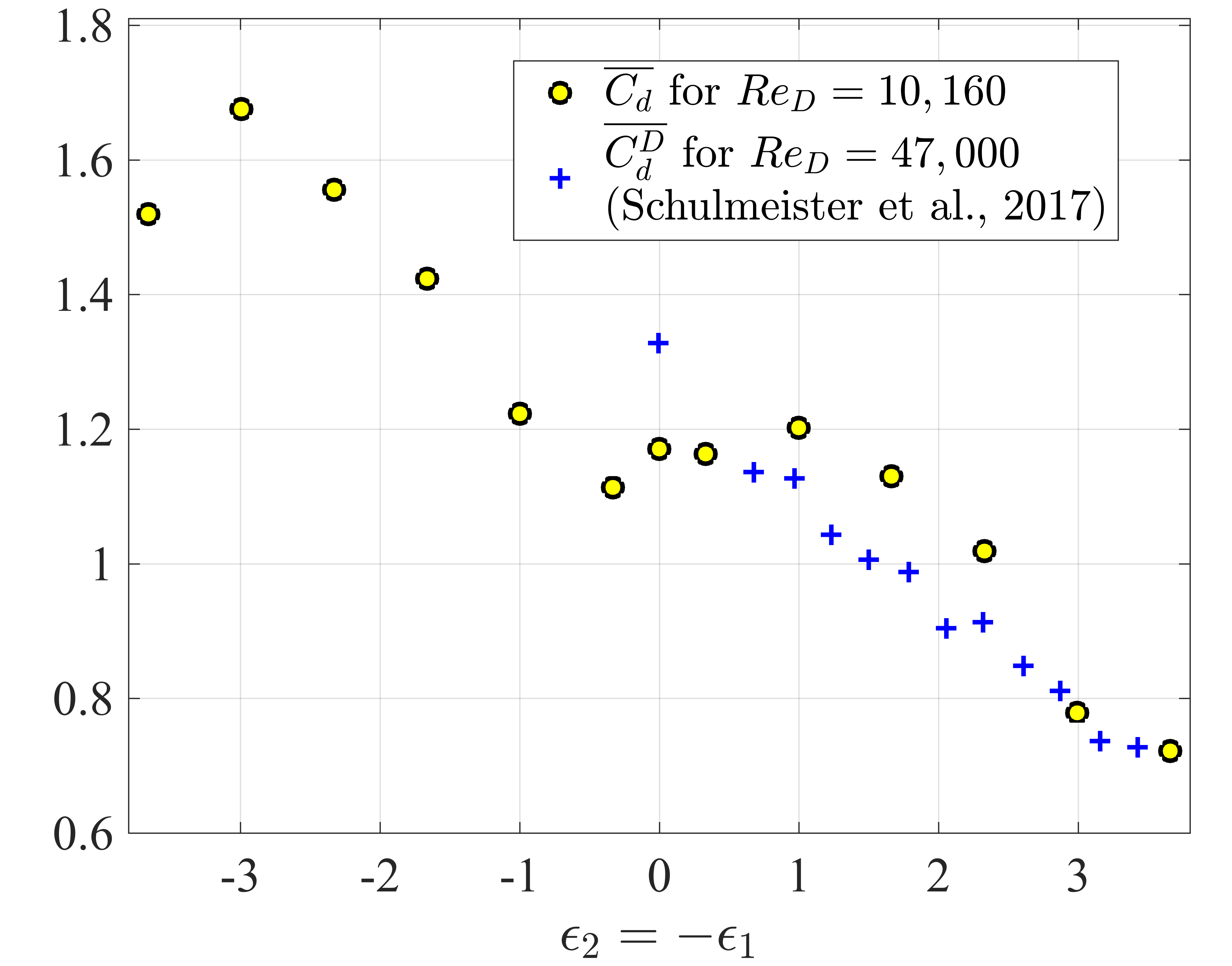}
\subcaption{}
\end{subfigure}
\caption{Hydrodynamic coefficients versus rotation rates $\epsilon_1$ and $\epsilon_2$ at $Re_D = 10,160$: (a) $\overline{C_d}$, (b) $\overline{C_l}$, and (c) comparison of $\overline{C_d}$ in the current experiment with $\overline{C_d^D}$ in \cite{schulmeister2017flow}. The star in (a) represents $\overline{C_{d0}} = 1.171$ when $\epsilon_1 = \epsilon_2 = 0$, and the white dashed line in (b) represents the contour line of $\overline{C_l} = 0$.}
\label{fig:Uniform}
\end{figure}

We first conducted 169 experiments with the control cylinders rotating at a constant speed. The result of the average drag and lift coefficients $\overline{C_d}$ and the $\overline{C_l}$ is plotted in Figs. \ref{fig:Uniform} (a) and (b). The result shows that the $\overline{C_d}$ decreases from 1.7 to 0.72 as $\epsilon_1$ decreases and $\epsilon_2$ increases. This decrease was the effect of the fast rotating control cylinders that help reattach the previously separating boundary layer in the main cylinder, reducing the wake width and hence the pressure drag \citep{schulmeister2017flow}.

A comparison is made between the $\overline{C_d}$ in the current experiment at $Re_D = 10,160$ and the $\overline{C_d^D}$ in the experimental work by \cite{schulmeister2017flow} at $Re_D = 47,000$. Fig. \ref{fig:Uniform} (c) demonstrates the same trend of the mean drag coefficient against $\epsilon_2 = -\epsilon_1$ for both sets of results. Note that Fig. \ref{fig:Uniform} (b) shows that when $\epsilon_2 = -\epsilon_1$, $\overline{C_l}$ is close to zero.

The experiment comparison when the control cylinders rotate at a constant speed confirms the validity of the current experimental setup, and we find that the minimum of the $\overline{C_d}$ happens with $\epsilon_1$ at full speed in the clock-wise (CW) direction, while $\epsilon_2$ at full speed in the counter clock-wise (CCW) direction; $\overline{C_l}$ is found to be close to zero.

\subsection{Reinforcement learning in experimental environments}

\subsubsection{Task one: drag reduction}
\begin{figure}
\centering
\begin{subfigure}{0.32\columnwidth}
\centering
\includegraphics[width=\columnwidth,keepaspectratio]{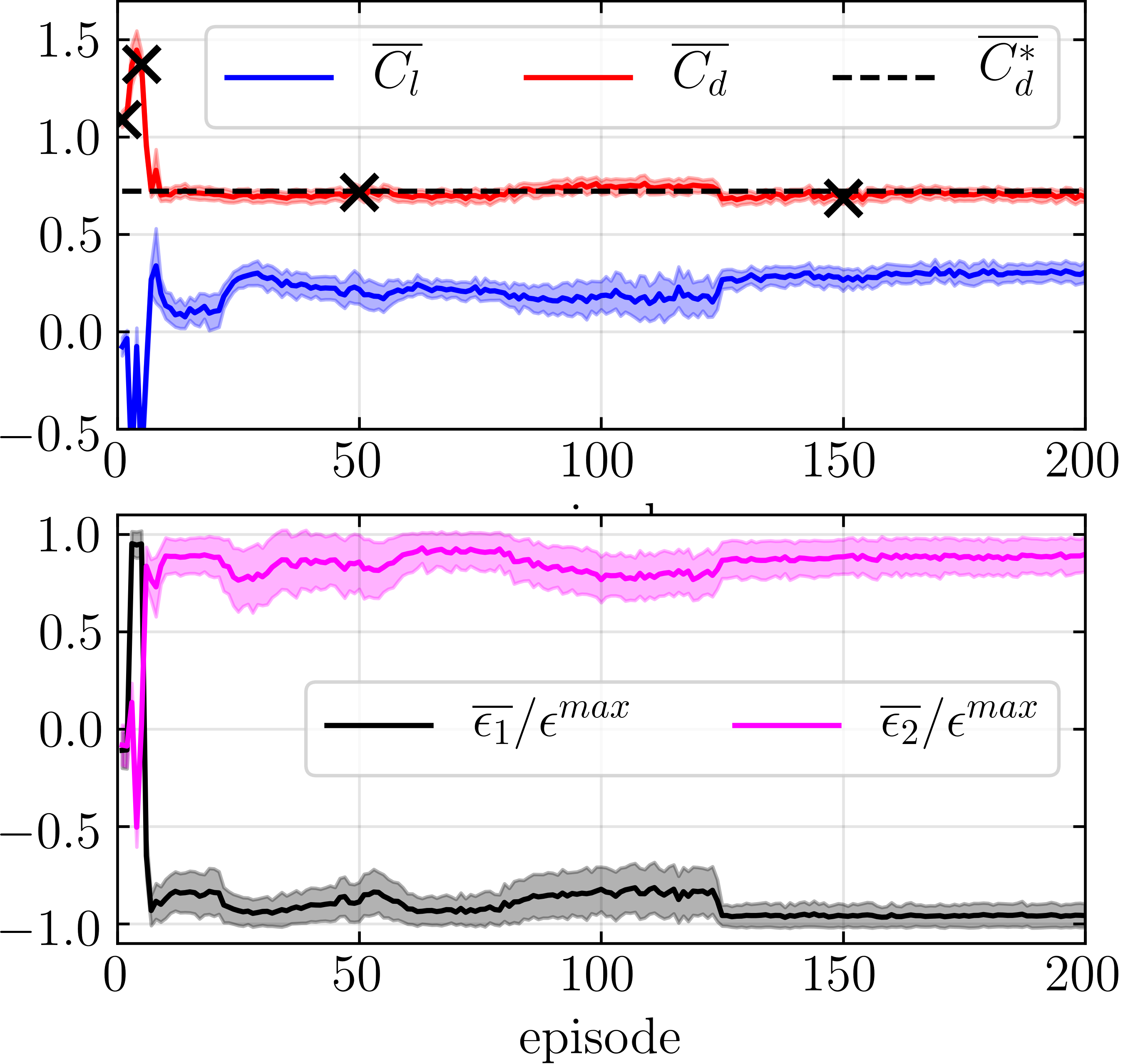}
\subcaption{}
\end{subfigure}
\begin{subfigure}{0.32\columnwidth}
\centering
\includegraphics[width=\columnwidth,keepaspectratio]{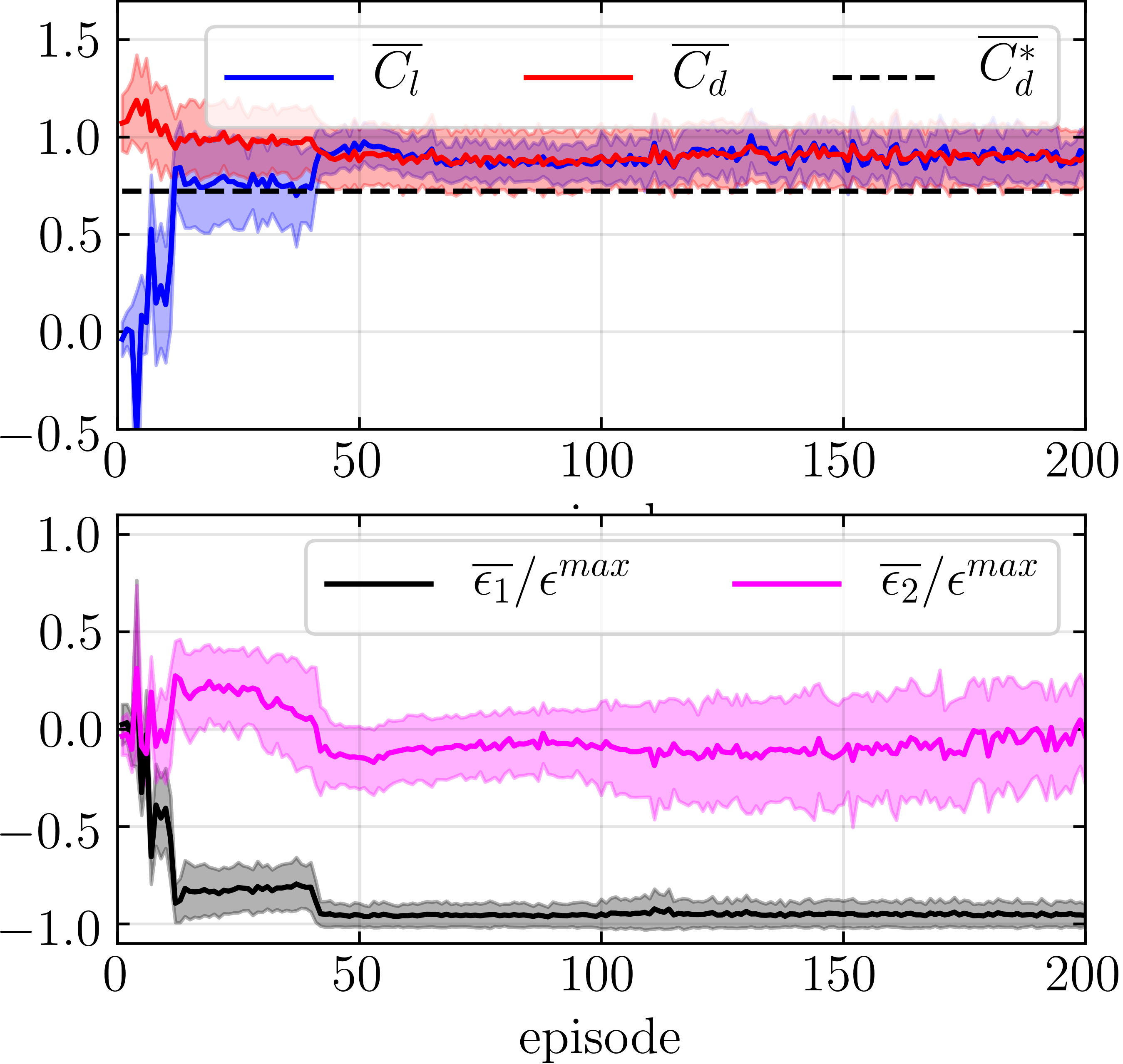}
\subcaption{}
\end{subfigure}
\begin{subfigure}{0.32\columnwidth}
\centering
\includegraphics[width=\columnwidth,keepaspectratio]{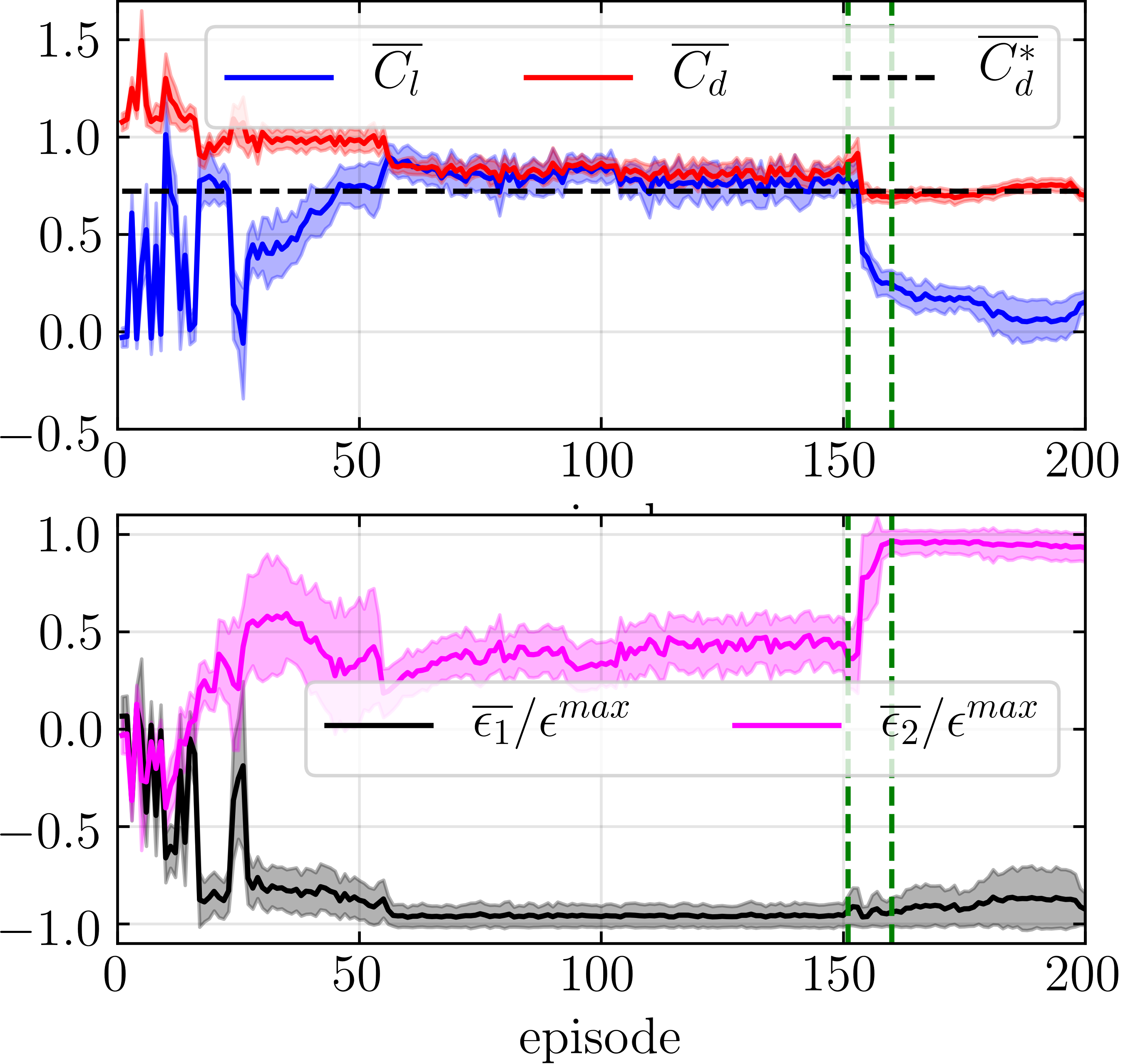}
\subcaption{}
\end{subfigure}
\begin{subfigure}{0.99\columnwidth}
\centering
\includegraphics[width=0.49\columnwidth,keepaspectratio]{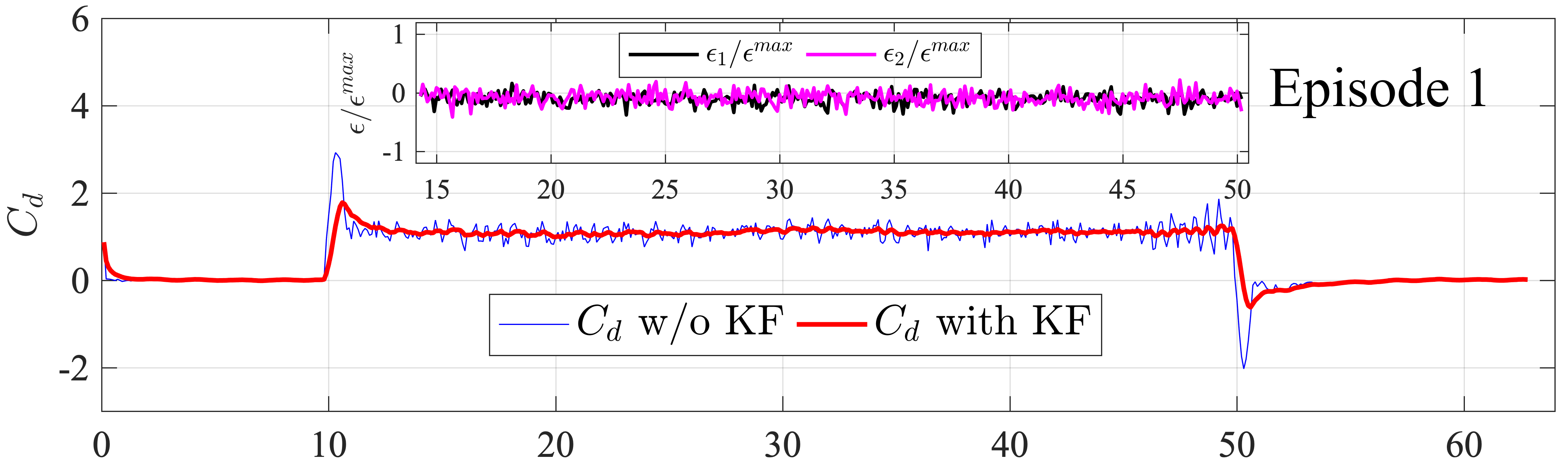}
\includegraphics[width=0.49\columnwidth,keepaspectratio]{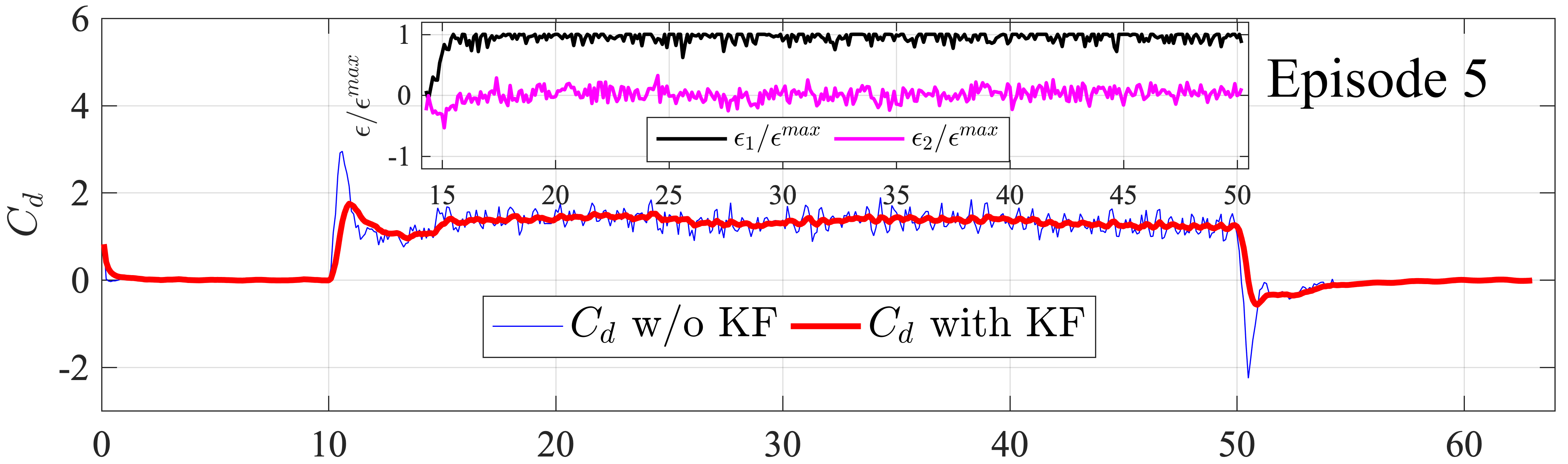}
\includegraphics[width=0.49\columnwidth,keepaspectratio]{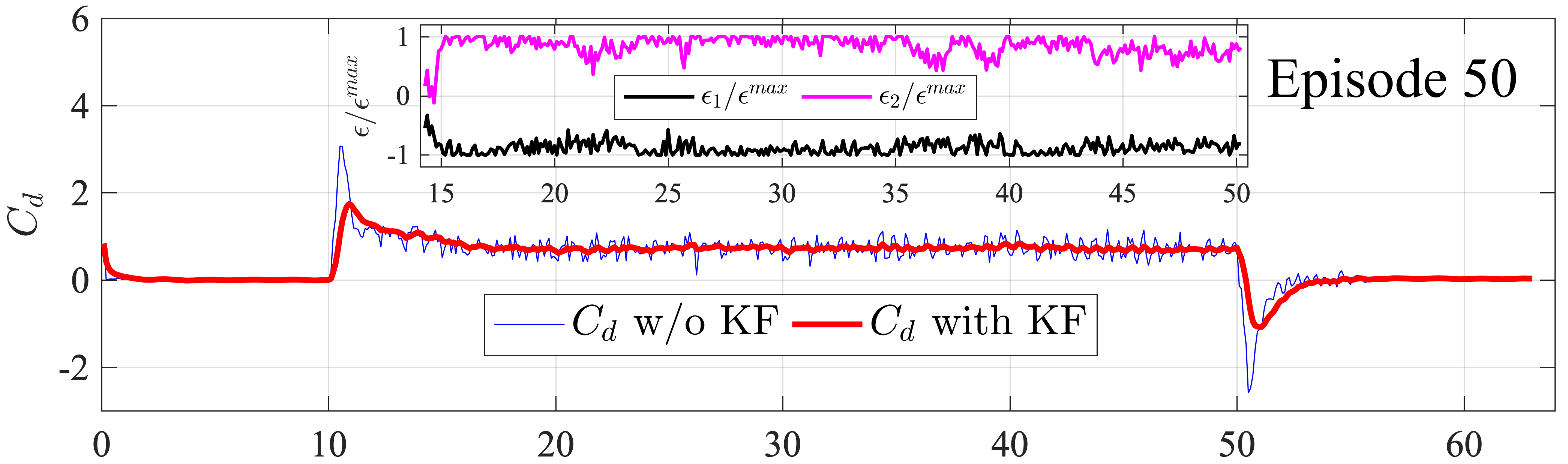}
\includegraphics[width=0.49\columnwidth,keepaspectratio]{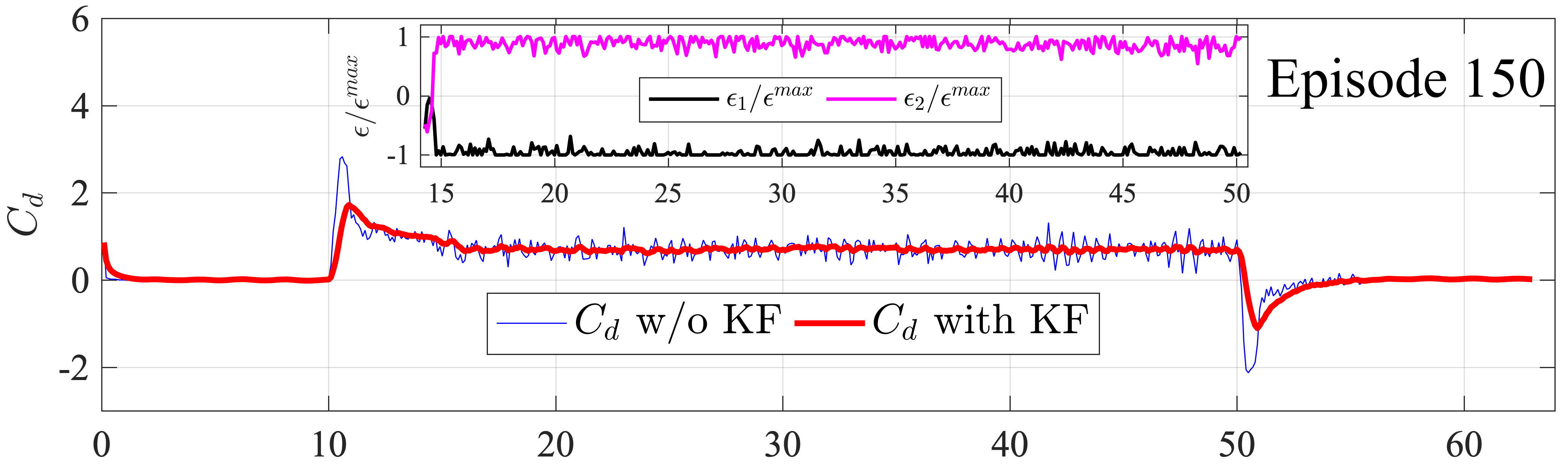}
\subcaption{}
\label{fig:EXPRLTime}
\end{subfigure}
\begin{subfigure}{0.99\columnwidth}
\centering
\includegraphics[width=\columnwidth,keepaspectratio]{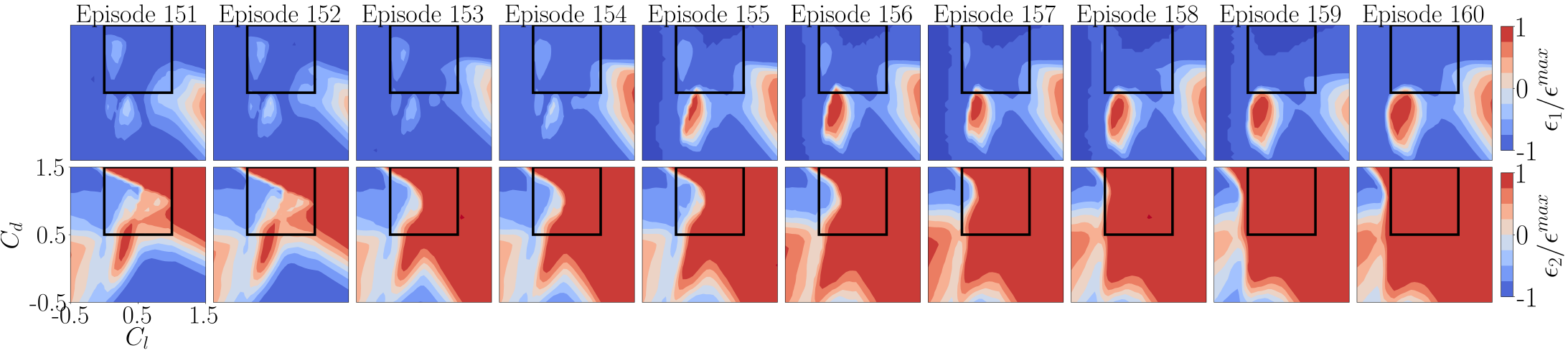}
\subcaption{}
\label{fig:policy}
\end{subfigure}
\caption{Training process in experimental task one. (a) case \Romannum{1}; (b) case \Romannum{2}; (c) case \Romannum{3}. The first row shows the hydrodynamic coefficients over 200 episodes, and the second row shows actions over 200 episodes. The solid lines and the shaded areas represent the mean value and one standard deviation over each episode, respectively. The black dashed lines represent $\overline{C^*_d}$. (d) Time trace of drag coefficient and actions (inset) for episode 1, 5, 50 and 150 in case \Romannum{1}. The carriage moves at the $10^{th}$ second and stops at the $50^{th}$ second. The active control is switched on at the $14^{th}$ second. (e) Visualization of policy evolution from episode 151 to 160 in case \Romannum{3}, corresponding to the region between the two green dashed lines in (c). In (e), the first row shows $\epsilon_1/\epsilon^{max}$ and the second row shows $\epsilon_2/\epsilon^{max}$, in terms of $C_d$ and $C_l$.}
\label{fig:EXPCurve}
\end{figure}

Three cases have been tested to demonstrate the importance of an appropriately designed RL reward and the application of a Kalman filter (KF) \citep{zarchan2013fundamentals} for noise reduction when the agent inquires the states. The results of the $\overline{C_d}$, $\overline{C_l}$ as well as $\overline{\epsilon_{1,2}}/\epsilon^{max}$ for the first 200 episodes are plotted in Fig. \ref{fig:EXPCurve}, while the setup of the reward and the filter in the three cases are listed as follows:
\begin{enumerate}
    \item Case \Romannum{1}: $r = -sgn(C_d)C_d^2 - 0.1{C_l}^2$ with KF;
    \item Case \Romannum{2}: $r = -sgn(C_d)C_d^2 - 0.1{C_l}^2$ without KF;
    \item Case \Romannum{3}: $r = -sgn(C_d)C_d^2$ with KF.
\end{enumerate}

The result of case \Romannum{1} is shown in Fig. \ref{fig:EXPCurve} (a), and the $\overline{C_d}$ is found to drop quickly and converges to approximately $\overline{C^*_d}$ in about 10 episodes, i.e., about half an hour in wall clock time, where $\overline{C^*_{d}}$ is the minimum value found in the reference experiment of the control cylinders rotating at $\epsilon_2 = -\epsilon_1 = 3.66$. The learning curve of actions shows that the agent learns to rotate the two cylinders in the opposite directions with near-maximum speeds. We observe that the $\overline{C_d}$ increases in the first few episodes before decreasing and converging, which is a result of agent's random exploration in the early stage of learning.

The time traces of $C_d$ and actions of the four different episodes in case \Romannum{1} (highlighted with cross markers in Fig. \ref{fig:EXPCurve} (a)) are displayed in Fig. \ref{fig:EXPRLTime}. A comparison between the raw data (blue) and the filtered data (red) reveals that the KF manages to remove the high frequency oscillations in $C_d$. Fig. \ref{fig:EXPRLTime} (a) of the first episode shows that when the learning process just begins, the RL agent fails to make any informed decision. In the fifth episode shown in Fig. \ref{fig:EXPRLTime} (b), the RL agent explores the rotation of the first control cylinder at its maximum speed in the CCW direction, which results in an increase of $\overline{C_d}$. After tens of policy updates, at the $50^{th}$ episode shown in Fig. \ref{fig:EXPRLTime} (c), when the active control is turned on, the RL agent manages to make the correct decision to rotate the control cylinders in the right direction and at the right speed, and therefore, reduce the $\overline{C_d}$. Comparing the actions of the $150^{th}$ episode in Fig. \ref{fig:EXPRLTime} (d) to those of the $50^{th}$ episode, we see that the actions are more stable with less variation.

In case \Romannum{2}, we use the same reward as in the case \Romannum{1}, but do not employ a KF. The result in Fig. \ref{fig:EXPCurve} (b) shows that after 200 episodes, the RL agent fails to reduce the $\overline{C_d}$ as effectively as in case \Romannum{1} where KF is employed. The learning curve of actions indicates that the RL agent is not able to learn an appropriate policy for the second rotational cylinder, resulting in large $\overline{C_d}$ and $\overline{C_l}$. The comparison between case \Romannum{1} and \Romannum{2} clearly shows the necessity of noise reduction when applying RL techniques in experimental environments and real-world applications.

To demonstrate the importance of properly designing the reward, in case \Romannum{3}, we keep the KF implementation but change the reward to $-sgn(C_d)C_d^2$. The augmentation of the weighted squared lift coefficient in case \Romannum{1} is motivated by the need to reduce the oscillating lift force, through preventing the alternate shedding of a vortex street. The result shows that the $\overline{C_d}$ is reduced slowly over the number of episodes. Between the $55^{th}$ and the $150^{th}$ episodes, the $\overline{C_d}$ reaches a relatively constant value of about 0.83, higher than the $\overline{C^*_{d}} = 0.72$, and the $\overline{C_l}$ is as large as 0.77. With the increase of episodes, we observe tha at around the $155^{th}$ episode, the $\overline{C_d}$ drops suddenly and converges to the $\overline{C^*_{d}}$, while the magnitude of $\overline{C_l}$ decreases to a value close to zero. 

In order to explain such a drastic change of the hydrodynamic coefficients at around the $155^{th}$ episode, in Fig.~\ref{fig:EXPCurve} we visualize the policy evolution between the $151^{st}$ and the $160^{th}$ episodes: the RL agent policy is initially stuck at a local minimum but then manages to escape due to exploration. The $C_d$ and $C_l$ in the whole learning process are mostly concentrated in the region highlighted by the black square. Note that the policy for $\epsilon_2$ gradually approaches to the strategy of rotating with maximum speed, showing the process of learning in the time interval. In addition, the policy could be far from optimal outside the highlighted region, as the agent learns from the experience collected, and can hardly generalize the policy for outlier states.

\subsubsection{Task two: maximization of the system power gain efficiency}

\begin{figure}
\centering
\begin{subfigure}{0.51\columnwidth}
\centering
\includegraphics[width=\columnwidth,keepaspectratio]{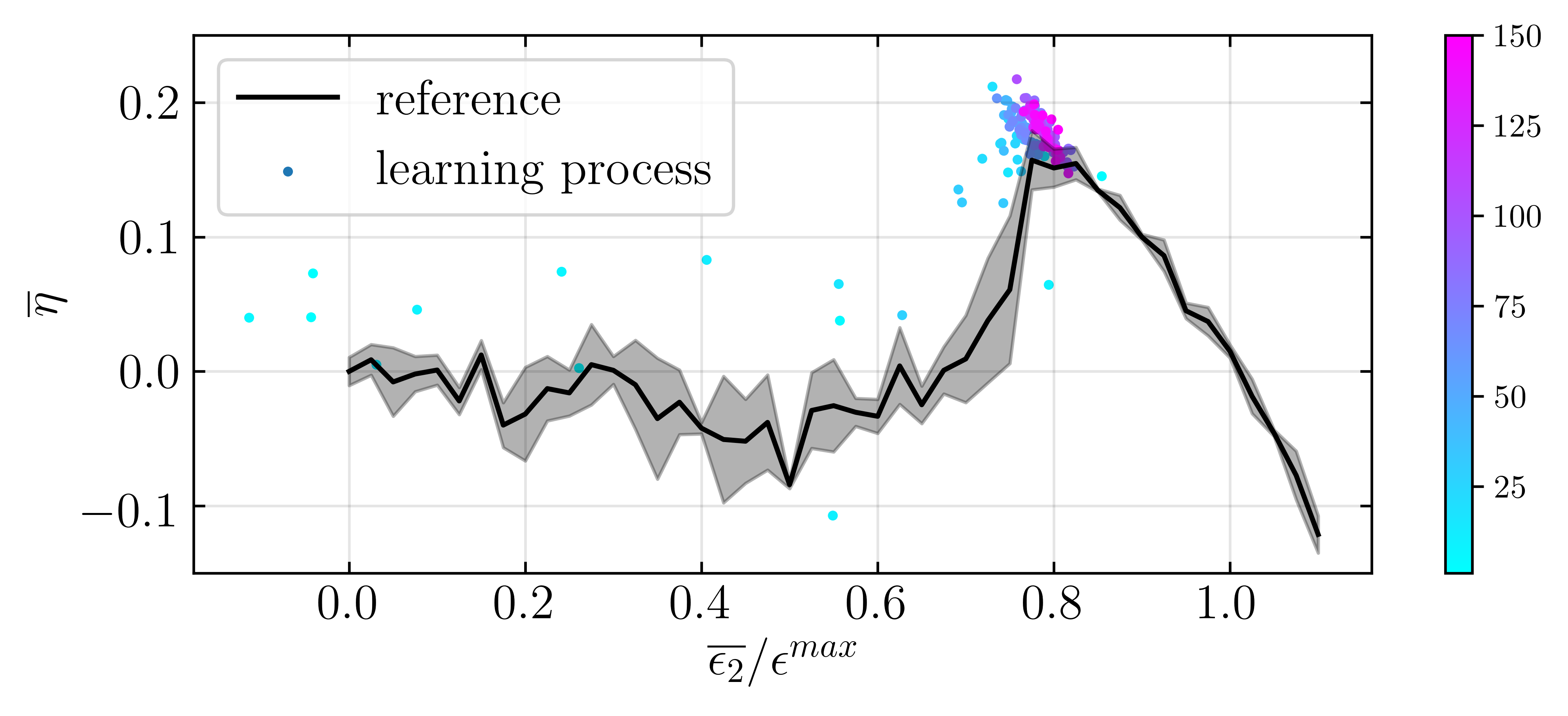}
\subcaption{}
\label{fig:Evol_2}
\end{subfigure}
\begin{subfigure}{0.45\columnwidth}
\centering
\includegraphics[width=\columnwidth,keepaspectratio]{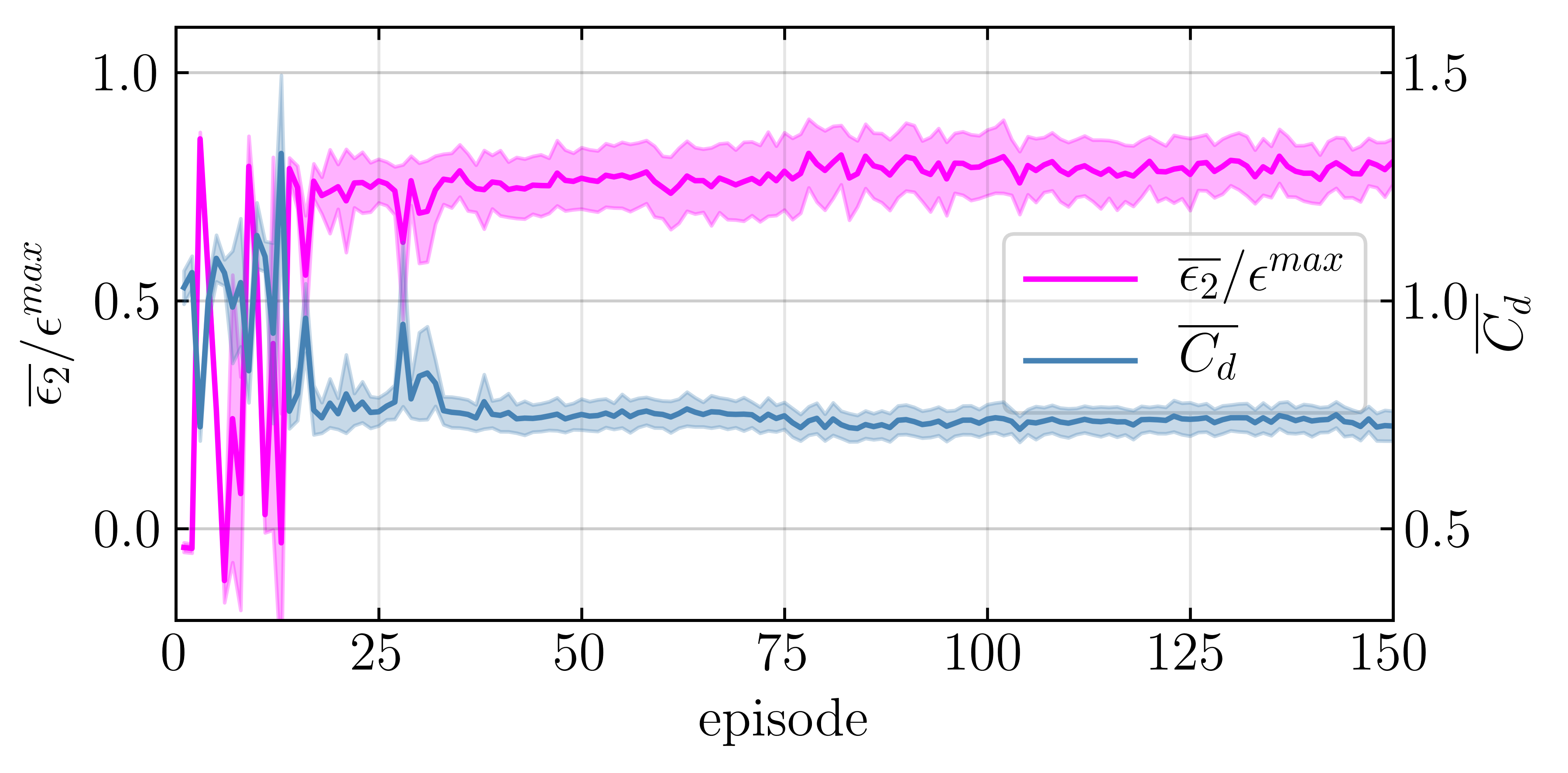}
\subcaption{}
\label{fig:Evol_1}
\end{subfigure}
\caption{Training process in experimental task two. (a) Evolution of average power gain efficiency $\overline{\eta}$ during the training process. The black solid line and the shaded area represent the mean and one standard deviation of $\overline{\eta}$ with constant $\epsilon_1 = - \epsilon_2$ over three independent experiments. The dots represent $\overline{\eta}$ in each episode, while the different colors represent different episode indices. (b) $\overline{\epsilon_2}$ and $\overline{C_d}$ in each episode.  The solid lines represent the mean value over each episode, while the shaded areas represent one standard deviation over each episode.}
\label{fig:Evol}
\end{figure}

We define the system power gain efficiency as $\eta = \Delta P/(0.5\rho U^3 DL)$, which is increased by the drag reduction, $\overline{C_{d0}} - C_d$, and decreased due to the power loss from the friction of the control cylinder rotation, $C_f\frac{\pi d}{D}(|\epsilon_1|^3 + |\epsilon_2|^3)$, where we restrict $\epsilon_1 = - \epsilon_2$ in this task, $\overline{C_{d0}}$ is the average drag coefficient when $\epsilon_1 = \epsilon_2 = 0$, and the friction coefficient is calculated as $C_f = 0.027/Re_d^{(1/7)} = 0.0097$ \citep{prandtl1949report}. Our goal is to maximize the \textit{average} system power gain efficiency $\overline{\eta}$ over one episode. Therefore, we constructed the reward function as follows,
\begin{equation}
    r = \eta = \frac{\Delta P}{0.5\rho U^3 DL} = [\overline{C_{d0}} - C_d] - [C_f \frac{\pi d}{D}(|\epsilon_1|^3 + |\epsilon_2|^3)].
\end{equation}

Due to the trade-off between the drag reduction and the power loss of the cylinder rotation, in this task, for the static control, the maximum of $\overline{\eta}$ is achieved at $\epsilon_2/\epsilon^{max} \approx 0.8$, shown in Fig.~\ref{fig:Evol}(a) by the black solid line as the reference. The dots in Fig.~\ref{fig:Evol}(a) represent $\overline{\eta}$ estimated in each episode, and are shown to be concentrated near the peak of the reference line for the episodes with well-trained RL agent. In fact, the optimal $\overline{\eta}$ from the RL experiment is found to be higher than the maximum from the static control, which could be explained by the control strategy designed by the agent that is dynamic instead of static. We also plot the $\overline{C_d}$ and the $\overline{\epsilon_2}/\epsilon^{max}$ for each episode in Fig~\ref{fig:Evol}(b).

\subsection{Reinforcement learning in simulation environments}

\begin{figure}
\centering
\begin{minipage}[b]{0.43\columnwidth}
    \centering
    \includegraphics[width=0.97\columnwidth,keepaspectratio]{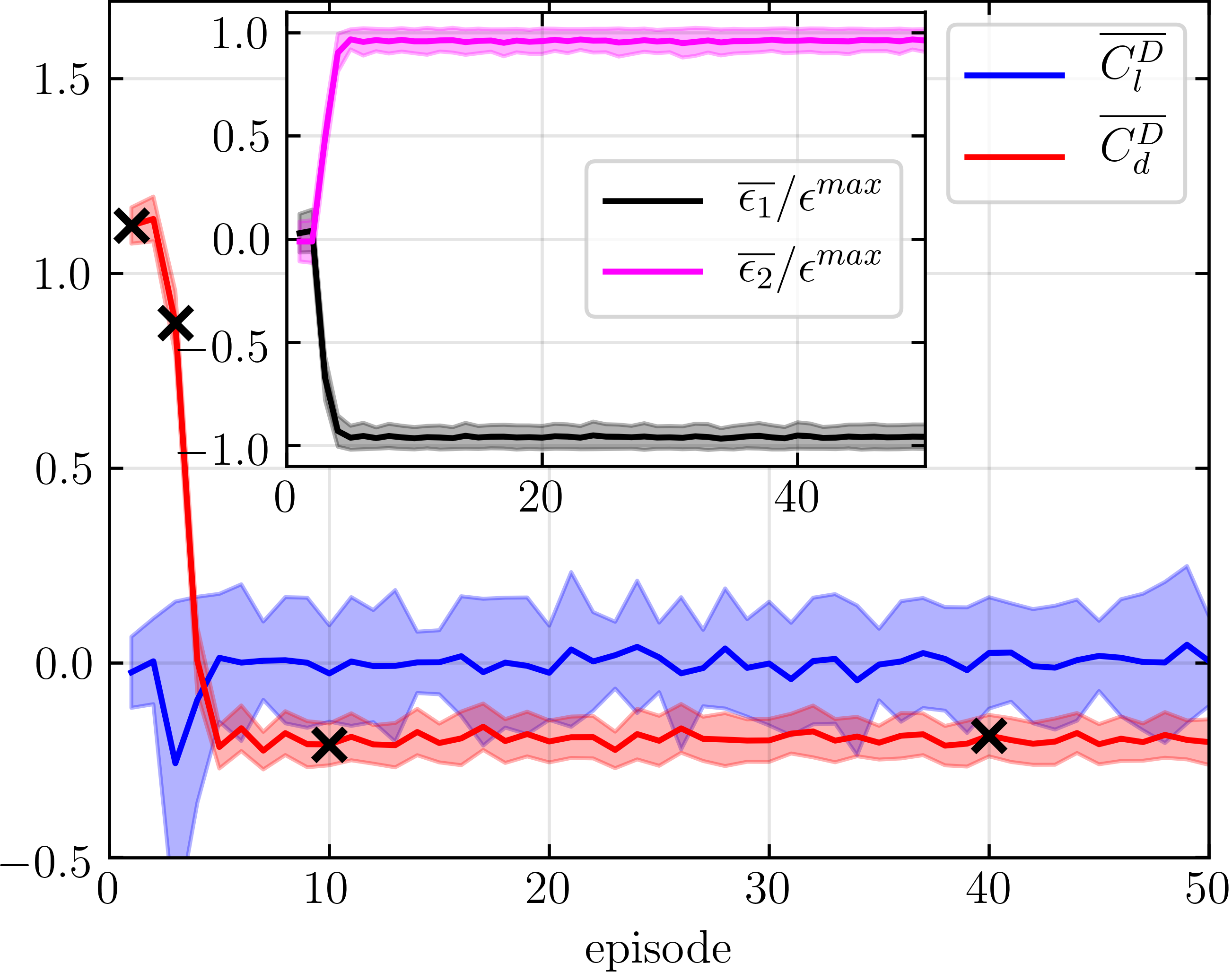}
    \subcaption{}
\end{minipage}
\begin{minipage}[b]{0.56\columnwidth}
    \includegraphics[width=0.49\columnwidth,keepaspectratio]{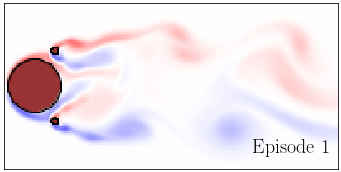}
    \includegraphics[width=0.49\columnwidth,keepaspectratio]{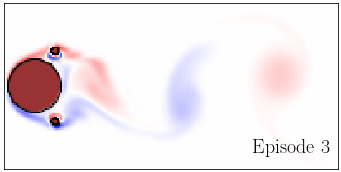}
    \includegraphics[width=0.49\columnwidth,keepaspectratio]{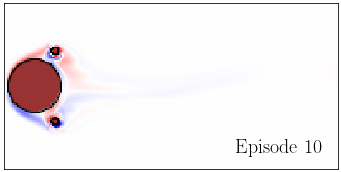}
    \includegraphics[width=0.49\columnwidth,keepaspectratio]{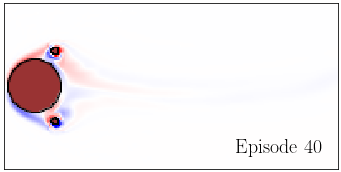}
    
    \hspace{0.1\columnwidth}
    \includegraphics[width=0.8\columnwidth,keepaspectratio]{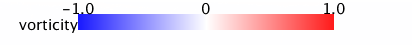}
     \subcaption{}
\end{minipage}
\caption{Training process in the simulation task. (a): Hydrodynamic coefficients and actions over 50 episodes in the simulation environment. The solid lines represent the mean value over each episode, while the shaded areas represent one standard deviation over each episode. (b): Visualization of the flow field in the end of episode 1, 3, 10 and 40.}
\label{fig:CFD}
\end{figure}

We conducted the drag reduction task in the simulation environment, where the parameters used are shown in Table \ref{tab:para}, and the results are displayed in Fig. \ref{fig:CFD} for a total of 50 episodes. The results demonstrate that only after four episodes the RL agent has already achieved a stable hydrodynamic performance, with the mean drag coefficient of the main cylinder $\overline{C_d^D}$ negative, and the mean lift coefficient $\overline{C_l^D}$ close to zero when the control cylinders rotate close to their maximum speeds and in opposite directions.

We select the first, the third, the tenth and the fortieth episode to visualize the flow behind the main and control cylinders. From Fig. \ref{fig:CFD}, we can see that at the first episode, the vortex shedding behind the main cylinder is strong and the width of the wake is wide, so the pressure drag is large. At the third episode, the rotation of the control cylinders results in a narrow width of the wake and a regular shedding vortex street, which leads to a smaller drag coefficient but enhanced oscillation of the unsteady lift coefficient term. At the tenth episode, the RL agent has learnt an appropriate control strategy: the maximum rotation speed of the two control cylinders results in a thrust force on the main cylinder as well as zero mean lift coefficient as they are able to stabilize the flow and eliminate the shed vorticies in the wake \citep{zhu2015simultaneous}. The similarity of the wake pattern and the hydrodynamic coefficients value between episode fortieth and tenth demonstrates the convergence and stability of the RL agent's active control strategy for the current system.

\section{Summary}
\label{secConc}
We demonstrated the feasibility of applying reinforcement learning with proper designed reward function to discover effective active control strategies in experimental environments, by studying the bluff body flow control problem of actively reducing the drag force and maximizing the system power gain efficiency through the use of two rotating smaller control cylinders attached to the main cylinder. With a properly designed reward function, the agent was able to learn a control strategy that is comparable to the optimal one found in static control experiments, within tens of experiments, requiring only several hours of wall-clock time.

The vortex shedding behind the cylinder was effectively suppressed by the control strategy learned by the agent, as was illustrated in the companion simulation studies. We also demonstrated the necessity of noise reduction techniques using a Kalman filter, a method which is especially suitable for experimental setups.

We believe that the flow control problem studied in the current work is only the beginning, and reinforcement learning can find wide applicability in a variety of experimental fluid mechanics problems, especially high-dimensional dynamic fluid control problems that are too difficult to tackle with classical methods. The learning algorithm we employ is totally model-free, but it will be worth exploring the possibility of incorporating some domain knowledge and designing physics-informed reinforcement learning algorithms, in order to further accelerate the scientific discovery.

The code for the reinforcement learning agent and simulation environment is shared and can be downloaded via the link of \url{https://github.com/LiuYangMage/RLFluidControl}.

\section*{Acknowledgement}
DF and MST would like to acknowledge support from the MIT Riser Digital Twin Consortium, and a fellowship provided by Shell International Exploration and Production, inc; YL and GEK would like to acknowledge support by DOE PhILMs project (No. DE-SC0019453).

\appendix

\section{Experimental model}
\label{sec:appA}

\begin{figure}
\centering
\includegraphics[width=0.6\columnwidth,keepaspectratio]{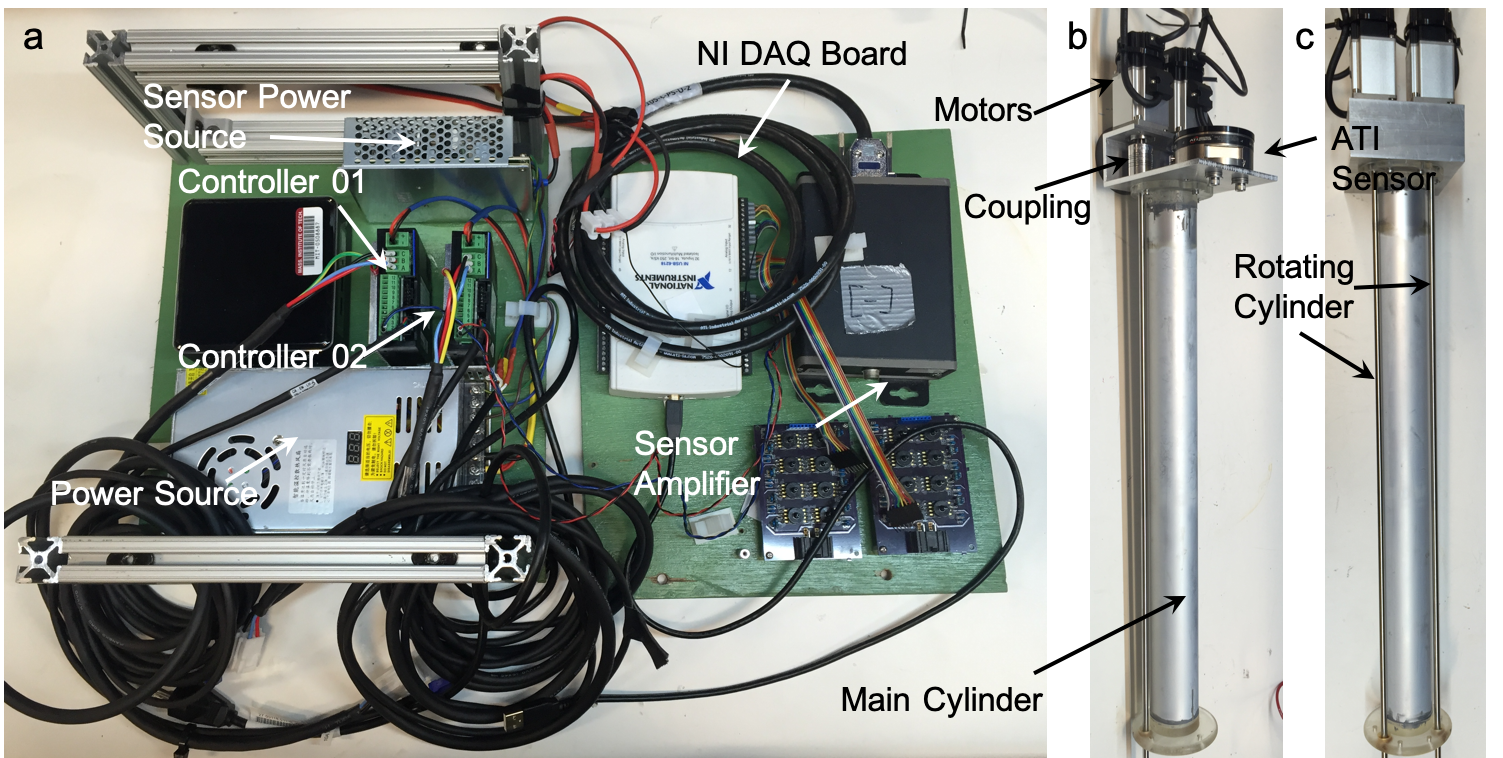}
\caption{Images of the control panels (a) and model (b: side view; c: back view) used in the experiment.}
\label{fig:RLSystem}
\end{figure}

In Fig. \ref{fig:RLSystem} we show the control panels and model used in the experiment. The control panels in Fig. \ref{fig:RLSystem} (a) have two decks and consist of six major components: two DYN2-series motor controllers, one NI USB-6218 Data Acquisition (DAQ) board, one ATI sensor amplifier, and two power sources. The DAQ board in the upper deck is controlled through the USB communication, and it is in charge of the analog data collection from the sensor amplifier at a sampling rate of 1000 Hz, and sending the signal to the two motor controllers at a feedback rate of 10 Hz. In the lower deck, two independent DYN2-series servo motor controllers for two DST-410 servo motors are powered by 60V DC power source.

Images of the experimental model are shown in Fig. \ref{fig:RLSystem} (b) and (c) for two views. The main cylinder is made from hard-anodized aluminium tube to prevent corrosion in the water. The two smaller stainless steel cylinders are connected to the two DST-410 motors via couplings and are supported by the bearings on both ends. Shown in Fig. \ref{fig:RLSystem} (b), the ATI-Gamma sensor is installed on top of the model and is used to measure the total lift and drag force on the main and two smaller cylinders altogether.

\bibliographystyle{unsrt}
\bibliography{RotCylRL.bib}

\end{document}